\documentclass[aps,pre,twocolumn,showpacs]{revtex4}
\usepackage{graphicx}
\usepackage{amsmath,amscd,amssymb}
\usepackage{color}

\newcommand{\be}{\begin{equation}}
\newcommand{\ee}{\end{equation}}
\newcommand{\bea}{\begin{eqnarray}}
\newcommand{\eea}{\end{eqnarray}}
\newcommand{\lan}{\left\langle}
\newcommand{\ran}{\right\rangle}
\newcommand{\br}{\mathbf{r}}

\newcommand{\bp}{\mathbf{p}}

\newcommand{\bom}{\mathbf{\Omega}}

\newcommand{\e}{\varepsilon}

\newcommand{\te}{\tilde{\epsilon}}

\begin{document}

\title{Dipolar depletion effect on the differential capacitance of carbon based materials}

\author{Sahin Buyukdagli$^{1}$\footnote{email:~\texttt{sahin\_buyukdagli@yahoo.fr}} and T. Ala-Nissila$^{1,2}$\footnote{email:~\texttt{Tapio.Ala-Nissila@aalto.fi}}}
\affiliation{$^{1}$Department of Applied Physics and COMP center of Excellence, Aalto University School of Science, P.O. Box 11000, FI-00076 Aalto, Espoo, Finland\\
$^{2}$Department of Physics, Brown University, Providence, Box 1843, RI 02912-1843, U.S.A.}
\date{\today}

\begin{abstract}
The remarkably low experimental values of the capacitance data of carbon based materials in contact with water solvent needs to be explained from a microscopic theory in order to optimize the efficiency of these materials. We show that this experimental result can be explained by the dielectric screening deficiency of the electrostatic potential, which in turn results from the interfacial solvent depletion effect driven by image dipole interactions. We show this by deriving from the microscopic system Hamiltonian a non-mean-field dipolar Poisson-Boltzmann equation. This can account for the interaction of solvent molecules with their electrostatic image resulting from the dielectric discontinuity between the solvent medium and the substrate. The predictions of the extended dipolar Poisson-Boltzmann equation for the differential capacitance are compared with experimental data and good agreement is found without any fitting parameters.
\end{abstract}

\maketitle

\section{Introduction}

New generation supercapacitors are used for a broad range of applications in nanoscopic scale technologies. In water purification technology, capacitive desalination is an efficient candidate that might replace the current leading technics such as reverse osmosis, a membrane based purification method known to suffer from the membrane fouling phenomenon~\cite{rev1}. Supercapacitors are also used as low cost and long life energy storage devices with considerably higher energy densities than conventional electrolytic capacitors~\cite{rev2}. A through understanding of the double layer structure of these devices is thus necessary to optimize their efficiency.

The understanding of the double layer structure was limited for several decades to the Gouy-Chapman-Stern model~\cite{rev3}. This model was later completed by considering additional effects specific to electrolyte solutions, to name but a few, the steric layer associated with the size of solvent molecules as well as the dipolar alignment close to the interface~\cite{Blum}, non-local effects in electrolytes at metallic interfaces~\cite{Kor1}, ionic crowding~\cite{Kor2} and overscreening~\cite{Kor3}.

Supercapacitors are commonly fabricated from carbon based materials with a dielectric permittivity $\e_m\approx 2-5$ much lower than the permittivity of the water solvent $\e_w=78$. The polarization of the interface resulting from this large dielectric discontinuity can drastically change the physics of the double layer. Image dipole interactions were considered in Ref.~\cite{mac} for a metallic interface. However, the work accounted exclusively for the effect of image interactions on the dipolar orientation without considering their role on the interfacial dipole density. Furthermore, it was recently shown in Ref.~\cite{prlnetz} that the Gouy-Chapman (GC) capacitance largely overestimates the experimental data obtained for carbon based materials. The failure of the GC capacitance was explained by the unability of the Poisson Boltzmann formalism to account for non-local dielectric effects.

In order to gain insight about the contribution of surface polarization effects on the capacitance of low dielectric substrates, we introduce in this work a first microscopic modeling of solvent molecules beyond the MF level approximation. Namely, we derive an extended dipolar PB (EDPB) equation that can self-consistently take into account the interfacial solvent depletion. This depletion results from the interaction of solvent molecules (modeled as dipoles) with their electrostatic image, an effect absent in the mean-field level DPB equation~\cite{dundip,prlorl}. The prediction of the EDPB equation is shown to agree well with experimental data for the differential capacitance of carbon based materials. However, it is also shown that the DPB formalism yields the same result as the PB equation, that is, it overestimates the experimental data by one order of magnitude. These observations strongly suggests that the dielectric discontinuity between the substrate and the solvent can solely explain the observed low values of the differential capacitance of carbon based materials, unlike the conclusion of Ref.~\cite{prlnetz} where it was argued that the surface polarity does not play a major role in the differential capacitance. Our results are also in agreement with the experimental work in Ref.~\cite{exp}, where the surface hydrophobicity was actually shown to strongly reduce the capacitance of carbone nanotubes.

\section{Extended Dipolar Poisson Boltzmann (EDPB) equation}

We will present in this part the derivation of an extended dipolar Poisson-Boltzmann formalism. The field-theoretic partition function of ions immersed in a dipolar liquid was derived in Ref.~\cite{dundip} as a functional integral over a fluctuating electrostatic potential $\phi(\br)$ in the form $\mathcal{Z}=\int \mathcal{D}\phi\;e^{-H[\phi]}$, where the Hamiltonian functional is given by
\bea\label{HamFunc2}
H[\phi]&=&\int\mathrm{d}\br\left[\frac{\left[\nabla\phi(\br)\right]^2}{8\pi\ell_B(\br)}-i\sigma(\br)\phi(\br)\right]\\
&&-\int\frac{\mathrm{d}\br\mathrm{d}\bom}{4\pi}\lambda_de^{E_d-V_w(\br)+i\left(\bp\cdot\nabla\phi\right)}\nonumber\\
&&-\sum_i\lambda_i \int\mathrm{d}\br e^{E_i-V_w(\br)+i \left[q_i\phi(\br)\right]}.\nonumber
\eea
The first integral term of the Hamiltonian~(\ref{HamFunc2}) is composed of the Maxwell tensor associated with a freely propagating electric field $\nabla\phi(\br)$ in the air, and a second part that couples the corresponding potential $\phi(\br)$ to a fixed surface charge distribution $\sigma(\br)$. The second and third integrals respectively account for the presence of solvent molecules (point dipoles) and ions of different species  denoted by the index $i$. Moreover, $\br=(x,y,z)$ is the configurational space and $\bom=(\theta,\varphi)$ stands for the solid angle characterizing the orientation of solvent molecules, with $\theta$ the angle between the dipole and the $z$ axis. We note that the external wall potential $V_w(\br)$ in Eq.~(\ref{HamFunc2}) restricts the space accessible to the particles, and in the case of the single dielectric interface located at $z=0$, it is of the form $V_w(z<0)=\infty$ and $V_w(z>0)=0$. Furthermore, $\lambda_d$ and $\lambda_i$ are respectively the fugacity of dipoles and ions, $\bp$ the dipole moment vector, and $q_i$ stands for the valency of ions for the species $i$. The heterogeneous Bjerrum length is defined as $\ell_B(\br)=e^2/\left[4\pi\e(\br)k_BT\right]$, where $e$ is the elementary charge, $T=300$ K is the ambient temperature, and $\e(\br)=\e_0\theta(z)+\e_m\theta(-z)$ is the dielectric permittivity of the medium in the absence of solvent molecules for the same single planar interface geometry. More precisely, $\e_0$ and $\e_m$ denote respectively the dielectric permittivity of the air (the subspace in $z>0$) and the low dielectric subbstrate located at $z<0$. From now on, the dielectric permittivities will be expressed in units of $\e_0$. We also note that the Bjerrum length in the air is $\ell_B\approx 55$ nm. Finally, the self energy of ions and polar molecules that are substracted from the potential and electrostatic field respectively read $E_i=\frac{q_i^2}{2}v_c^b(\br-\br')|_{\br=\br'}$ and $E_d=\frac{1}{2}(\bp\cdot\nabla_\br)(\bp\cdot\nabla_{\br'})v_c^b(\br-\br')$, where the Coulomb operator in the air is defined as ${v^b_c}^{-1}(\br,\br')=-\frac{k_BT\e_0}{e^2}\Delta\delta(\br-\br')$.

In this work, we aim at investigating the model of Eq.~(\ref{HamFunc2}) beyond the MF approximation where surface polarization effects are absent~\cite{dundip, prlorl}. One way to progress consists in opting for a variational minimization procedure that aims at finding the upper boundary for the dimensionless Grand potential of the system $\Omega=-\ln Z$ by minimizing the variational Grand potential defined as $\Omega_v=\Omega_0+\lan H-H_0\ran_0$, where the reference Hamiltonian is a Gaussian functional of the form
\be
H_0=\frac{1}{2}\int_{\br,\br'}\left[\phi(\br)-i\phi_0(\br)\right]
v^{-1}_0(\br,\br')\left[\phi(\br')-i\phi_0(\br')\right].
\label{H0phi}
\ee
Furthermore, $\phi_0(z)$ is a variational external potential and the electrostatic trial kernel is chosen in the same form as in Refs.~\cite{pre,prl,jcp},
\be\label{DH1}
v_0^{-1}(\br,\br')=\frac{k_BT}{e^2}\left[-\nabla(\e_v(\br)\nabla)+\e_v(\br)\kappa_c^2(\br)\right]\delta(\br-\br'),
\ee
where the piecewise variational dielectric permittivity is defined as $\e_v(\br)=\e_w\theta(z)+\e_m\theta(-z)$ and the trial screening length is given by $\kappa_c(\br)=\kappa_c\theta(z)$. After performing the functional integrals over $\phi(\br)$, one gets
\bea
\label{vargrpot}
\Omega_v&=&\Omega_{0}+\frac{k_BT}{2e^2}\int\mathrm{d}\br\mathrm{d}\br'\delta(\br-\br')\nonumber\\
&&\hspace{9mm}\times\left\{\left[\e(\br)-\e_v(\br)\right]\nabla_\br\cdot\nabla_{\br'}-\e_v(\br)\kappa_c^2(\br)\right\}v_0(\br,\br')\nonumber\\
&&+\int\mathrm{d}\br\left\{\sigma(\br)\phi_0(\br)-\frac{k_BT}{2e^2}\e(\br)\left[\nabla\phi_0(\br)\right]^2\right\}\\
&&-\sum_i\int\mathrm{d}\br\rho_i(\br)-\int\frac{\mathrm{d}\br\mathrm{d}\bom}{4\pi}\bar\rho_d(\br,\bom),\nonumber
\eea
where the gaussian contribution reads $\Omega_0=-\ln\int \mathcal{D}\phi\;e^{-H_0[\phi]}$. We also defined above the local ion density
\be\label{deni1}
\rho_i(\br)=\lambda_ie^{E_i-V_w(\br)}e^{-q_i\phi_0(\br)-\frac{q_i^2}{2}v_0(\br,\br)}
\ee
and the local density of dipoles with orientation $\bom$
\bea\label{dend1}
\bar\rho_d(\br,\bom)&=&\lambda_de^{E_d-V_w(\br)}\\
&&\times e^{-\bp\cdot\nabla\phi_0(\br)-\frac{1}{2}(\bp\cdot\nabla_\br)(\bp\cdot\nabla_{\br'})v_0(\br,\br')|_{\br'=\br}}.\nonumber
\eea

By taking the derivative of the variational Grand potential Eq.~(\ref{vargrpot}) with respect to $\kappa_c$ and $\e_v$, one gets $\kappa_c^2=4\pi\ell_w\sum_i\rho_{b,i}q_i^2$ and $\e_v=\ell_B/\ell_w=\e_w=1+\frac{4\pi}{3}\ell_Bp_0^2\rho_{bd}$. These two relations respectively introduce the Debye-Huckel screening parameter and the Debye-Langevin form for the bulk dielectric permittivity of the water medium $\e_w$. The additional variational equation for $\phi_0(z)$, i.e. $\delta\Omega_v/\delta\phi_0(\br)=0$ yields
\be
\label{varel}
\frac{\partial}{\partial z}\tilde\e(z)\frac{\partial\phi_0(z)}{\partial z}+4\pi\ell_B\sigma(z)+4\pi\ell_B\sum\rho_i(z)q_i=0,
\ee
where we took into account the translational symmetry of the electrostatic potential within the $(x,y)$ plane. We note that the variational minimization left us in  Eq.~(\ref{varel}) with a spatially varying dielectric permittivity of the form
\be\label{per1}
\tilde\e(z)=1-\frac{4\pi\ell_B}{\phi'_0(z)}\int\frac{\mathrm{d}\bom}{4\pi}\bar\rho_d(z,\bom)p_z,
\ee
where $p_z=p_0\cos\theta$ stands for the component of the dipolar moment vector $\bp$ in the $z$ direction. We will call Eq.~(\ref{varel}) the Extended Dipolar Poisson Boltzmann (EDPB) equation.

The fugacity of dipoles and ions can be related to their bulk density in the limit $z\to\infty$ of the equations~(\ref{deni1}) and~(\ref{dend1}).
By injecting the obtained relations  for the fugacities with the inverse of the kernel Eq.~(\ref{DH1})~\cite{pre} into Eqs.~(\ref{deni1}) and~(\ref{dend1}), the local density functions take the form
\bea\label{deni2}
&&\rho_i(z)=\rho_{b,i}e^{-V_w(z)}e^{-q_i\phi_0(z)-V_c(z)}\\
\label{dend2}
&&\bar\rho_d(z,\bom)=\rho_{bd}e^{-V_w(z)}e^{-\bp\cdot\nabla\phi_0(z)-V_d(z,\bom)},
\eea
where we defined the following ionic and dipolar potentials,
\bea
\label{ion3}
&&V_c(z)=\frac{q^2\ell_w}{2}\int_0^\infty\frac{\mathrm{d}kk}{\rho_c}\Delta e^{-2\rho_cz}\\
\label{dip3}
&&V_d(z,\bom)=U_d(z)+T_d(z)\cos^2\theta,
\eea
with the functions
\bea
\label{funu}
&&U_d(z)=\frac{\ell_wp_0^2}{4}\int\frac{\mathrm{d}kk^3}{\rho_c}\Delta e^{-2\rho_cz}\\
\label{funt}
&&T_d(z)=\frac{\ell_wp_0^2}{4}\int\frac{\mathrm{d}kk}{\rho_c}(2\rho_c^2-k^2)\Delta e^{-2\rho_cz}.
\eea
and $\Delta=(\rho_c-\eta k)/(\rho_c+\eta k)$, $\eta=\e_m/\e_w$, and $\rho_c=\sqrt{\kappa_c^2+k^2}$.
Carrying out the integral over $\theta$ in Eq.~(\ref{per1}) with the dipole density Eq.~(\ref{dend2}) and the potential Eq.~(\ref{dip3}), the local dielectric permittivity takes the form
\be\label{per2}
\tilde\e(z)=1+\frac{4\pi}{3}\ell_Bp_0^2\rho_{db}e^{-V_w(z)}e^{-U_d(z)}J(z),
\ee
where we defined the function
\bea
J(z)&=&\frac{3\sqrt\pi}{8T_d^{3/2}(z)}e^{\frac{p_0^2\phi'^2_0(z)}{4T_d(z)}}\left\{\mathrm{Erf}\left[\Psi_+(z)\right]+
\mathrm{Erf}\left[\Psi_-(z)\right]\right\}.\nonumber\\
&&-\frac{3e^{-T_d(z)}}{2T_d(z)}\frac{\sinh\left[p_0\phi'_0(z)\right]}{p_0\phi'_0(z)},
\eea
and the potentials
\bea
\Psi_\pm(z)=\frac{2T_d(z)\pm p_0\phi'_0(z)}{2\sqrt{T_d(z)}}.
\eea
The EDPB Eq.~(\ref{varel}) has to be solved numerically with  the ionic density profiles of Eq.~(\ref{deni2}) and the dielectric permittivity profile of Eq.~(\ref{per2}).

The second order differential equation~(\ref{varel}) should be solved with the boundary conditions $\phi_0(z\to\infty)=0$ and $\phi'_0(z\to 0^+)=2\e_w/\mu$, where the second boundary condition valid over the parameter domain $0\leq\e_m\leq\e_w$ follows by integrating Eq.~(\ref{varel}) in the close neighborhood of the interface, and noting that according to the dipolar potentials of Eqs.~(\ref{funu}) and~(\ref{funt}), one has $\rho_d(0)=0$ and $\te(0)=1$ on the boundary. We also note that in the limit where the potentials $V_c(z)$, $U_d(z)$, and $T_d(z)$ vanish, EDPB equation.~(\ref{varel}) reduces to the mean field DPB equation of Refs.~\cite{dundip,prlorl}.

We finally note that the orientation averaged density of solvent molecules is obtained according to $\rho_d(\br)=\int\frac{\mathrm{d}\bom}{4\pi}\bar\rho_d(\br,\bom)$. Evaluating the integral over $\theta$, the solvent density takes the form
\bea
\label{dend3}
\rho_d(z)&=&\rho_{bd}\frac{\sqrt\pi}{4\sqrt T_d(z)}e^{-V_w(z)}e^{-U_d(z)}e^{\frac{p_0^2\phi'^2_0(z)}{4T_d(z)}}\\
&&\times\left\{\mathrm{Erf}\left[\Psi_+(z)\right]+\mathrm{Erf}\left[\Psi_-(z)\right]\right\}.\nonumber
\eea

\section{Numerical results}

We will investigate in this part the EDPB Eq.~(\ref{varel}) for a symmetric electrolyte composed of two ion species of bulk densities $\rho_{b,i}=\rho_{bi}$ and valency $q_i=q$. All numerical results will be derived for monovalent ions ($q=1$) in contact with a negatively charged planar surface, i.e. $\sigma(z)=-\sigma_s\delta(z)$ with $\sigma_s>0$. We also note that within the convention adopted in this article, the surface charge $\sigma_s$ is expressed in units of the elementary charge $e$. Moreover, the model parameters $\rho_{db}$ and $\e_w$ are taken the same as in Ref.~\cite{prlnetz}. Namely, the bulk density of solvent molecules is $\rho_{db}=50.8$ M, which yields with the dipole moment $p_0=1$ {\AA} the bulk dielectric permittivity $\e_w=71$.

\begin{figure}
(a)\includegraphics[width=0.9\linewidth]{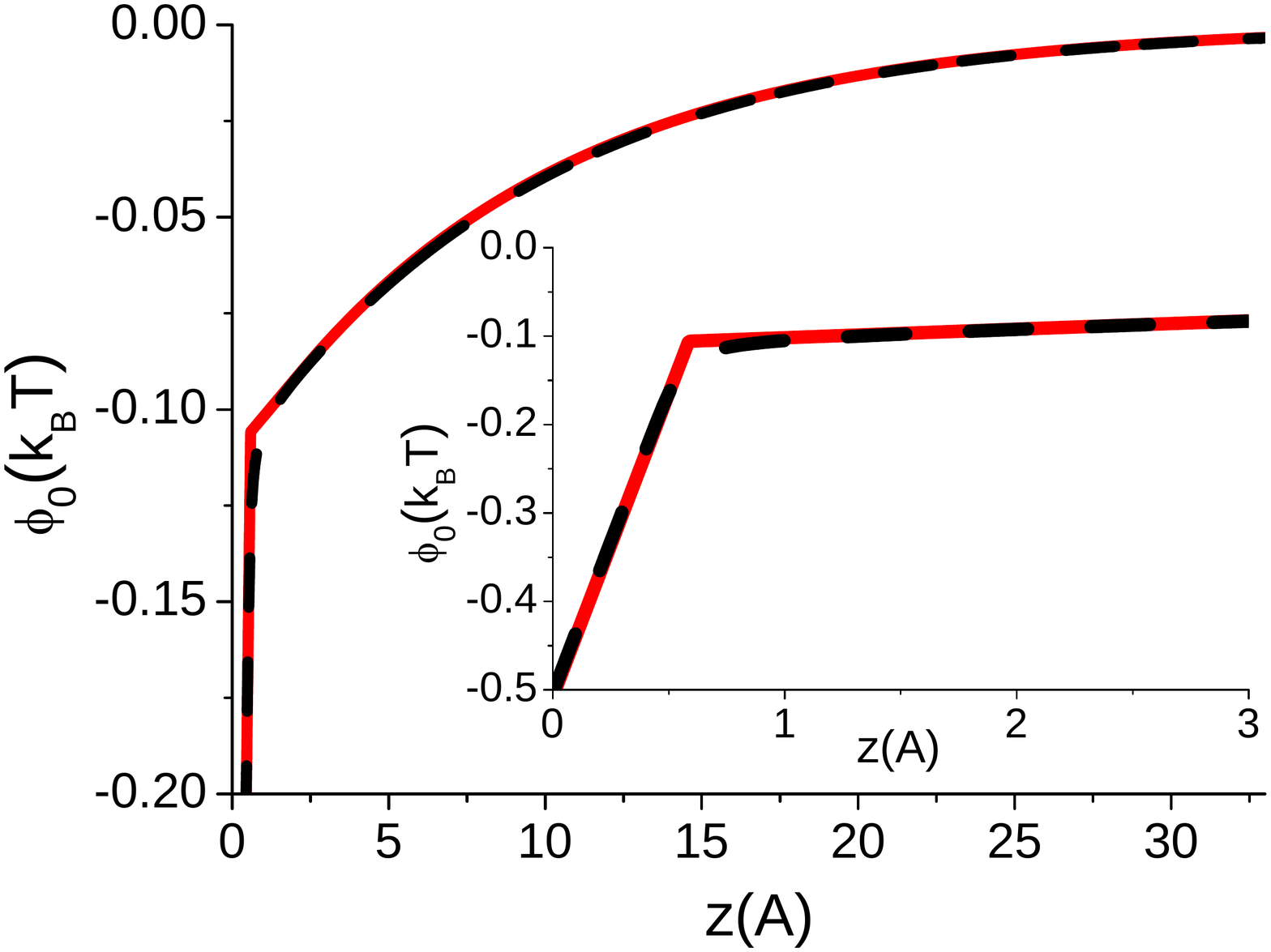}
(b)\includegraphics[width=0.9\linewidth]{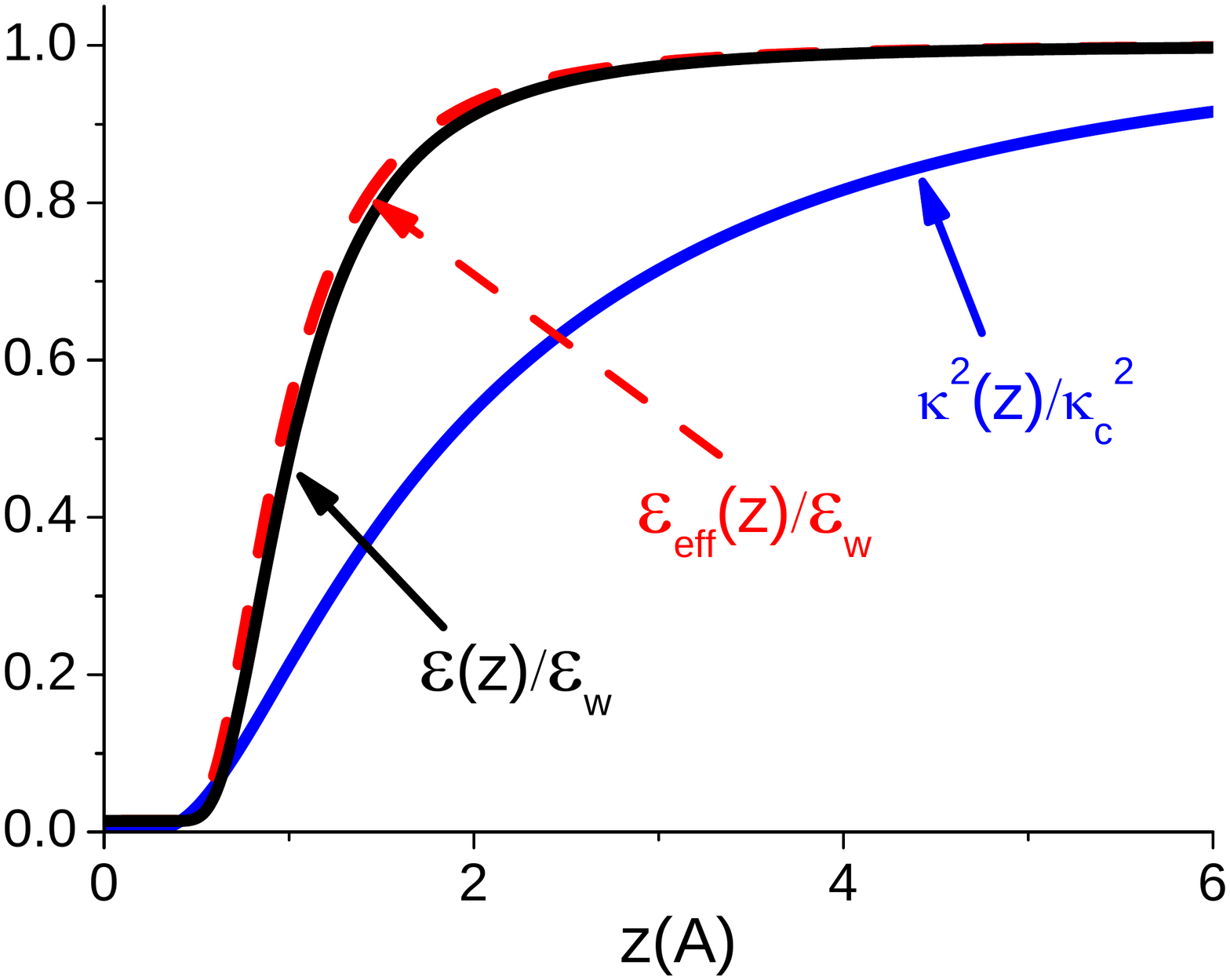}
\caption{(Color online)  (a) Electrostatic potential profile ($\sigma_s=0.01$ $\mbox{nm}^{-2}$) and (b) renormalized density and dielectric permittivity profiles for $\e_w=71$ and $\rho_{bi}=0.1$ M. The red line in (a) is from the restricted variational ansatz Eq.~(\ref{phires}) and the dashed black line corresponds to the solution of the EDPB equation.}
\label{den1}
\end{figure}

The potential profile obtained from the numerical solution of the EDPB Eq.~(\ref{varel}) for the parameters $\rho_{bi}=0.1$ M, $\e_m=1$ and $\sigma_s=0.01$ $\mbox{nm}^{-2}$ is reported in Fig.~\ref{den1}.a. One notices that the potential profile is composed of three regions, namely two successive layers close to the interface where $\phi_0(z)$ behaves as a linear function of $z$, and a third layer over which $\phi_0(z)$ exponentially decays.

In order to understand the form of the potential profile, we illustrate in Fig.~\ref{den1} the form of the dielectric permittivity $\te(z)$ and the screening parameter $\kappa^2(z)=\kappa_{c}^2e^{-V_c(z)}$ in the vanishing surface charge limit of Eq.~(\ref{varel}). It is seen that with increasing distance from the surface, the dielectric permittivity increases from the air permittivity $\te(z)=1$ to the bulk permittivity $\te(z)=\e_w$ over a distance  $h\approx 2$ {\AA}. We note that this dipolar exclusion effect is mainly due to the interaction of solvent molecules with their electrostatic images. Then, one sees that this solvent depletion regime is followed by an ionic depletion regime of thickness $d\approx 6$ {\AA}, an effect known to originate from image charge interactions~\cite{pre}.

Inspired by the behaviour of $\te(z)$ and $\kappa(z)$ that results from the interfacial depletion of solvent molecules and ions, we will introduce a restricted variational ansatz based on a piecewise trial solution for the electrostatic potential. We assume that $\phi_0(z)$ is the solution of Eq.~(\ref{varel}) in the linear limit of weak surface charge, with $\te(z)=\theta(h-z)+\e_w\theta(z-h)$ and $\kappa(z)=\kappa_c\theta(z-d)$, where the dipolar and ionic depletion lengths $h$ and $d$ are trial parameters that will be obtained from a numerical optimization procedure of the Grand potential of Eq.~(\ref{vargrpot}). The solution of Eq.~(\ref{varel}) with the above piecewise dielectric permittivity and ion density profiles, and satisfying the continuity of the potential $\phi_0(z)$ and the displacement field $D(z)=\te(z)\phi'_0(z)$ at $z=0$, $z=h$, and $z=d$, reads
\bea\label{phires}
 \phi_0(z)&=& -\frac{2}{\mu\kappa_c}\left[1+\kappa_c(d-h)\right]+\frac{2\e_w}{\mu}(z-h),\quad 0<z\leq h\nonumber\\
 \phi_0(z)&=&-\frac{2}{\mu\kappa_c}+\frac{2}{\mu}(z-d),\quad h\leq z\leq d\\
 \phi_0(z)&=&-\frac{2}{\mu\kappa_c}e^{-\kappa_c(z-d)},\quad z\geq d\nonumber.
\eea

Numerical optimization yields $h=0.6$ {\AA} and $d=2.3$ {\AA}. We note that due to the piecewise nature of the trial potential in Eqs.~(\ref{phires}), these values correspond approximately to half saturation densities. Figure~\ref{den1} shows that the potential profile obtained from the numerical optimization agrees very well with the general form obtained from the numerical solution of the EDPB equation. In Eqs.~(\ref{phires}), the first linear regime at $0<z\leq h$  corresponds to the solvent depletion layer resulting from image dipole interactions. This layer associated with dielectric screening deficiency is responsible for an amplification of the PB prediction of the surface potential by a factor of five. The second and third intervals correspond respectively to the usual ionic depletion and diffuse layers~\cite{pre}. The contribution of these layers to the differential capacitance will be investigated below.

\begin{figure}
(a)\includegraphics[width=0.9\linewidth]{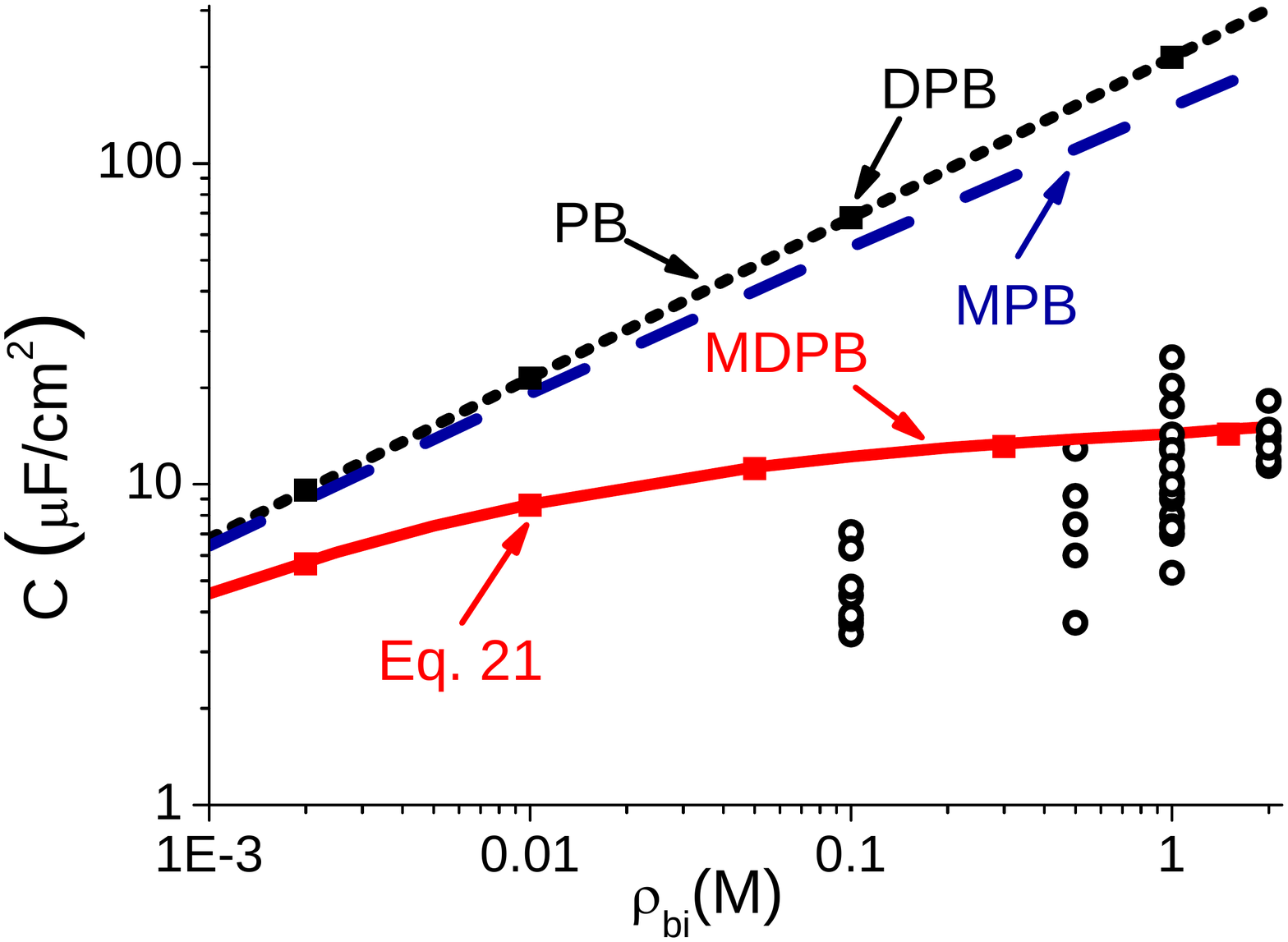}
(b)\includegraphics[width=0.9\linewidth]{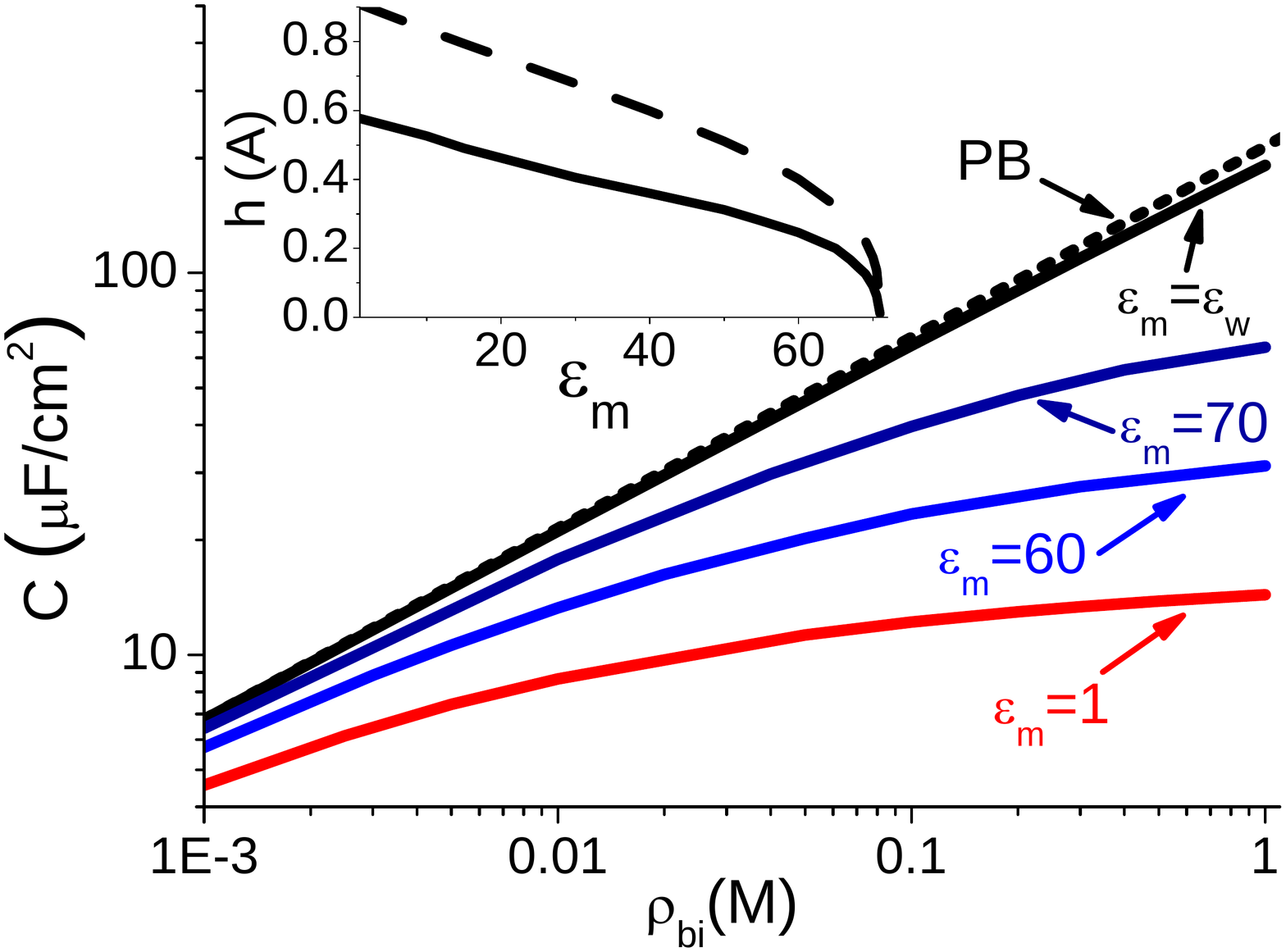}
\caption{(Color online)  (a) Differential capacitance against the bulk ion concentration for $\sigma_s=0$, $\e_m=1$, and $\e_w=71$. The black circles are the experimental data, the red solid curve is the result of the EDPB equation, the red squares are from Eq.~(\ref{cap2}), the dashed blue line is the MPB result, the dotted black line is the GC capacitance, and the black squares correspond to the prediction of the DPB equation. (b) The same plot as in (a) for various $\e_m$. The inset displays the evolution of the dipolar depletion length $h$ (solid curve) and $l_d$ (dashed curve) as a function of $\e_m$ for $\rho_{bi}=0.1$ M.}
\label{den2}
\end{figure}

The differential capacitance of the double layer is defined as
\be\label{cap1}
C_d=\frac{qe^2}{k_BT}\left|\frac{\partial\sigma_s}{\partial\phi_s}\right|,
\ee
where $\phi_s=\phi_0(z=0)$ is the surface potential. Fig.~\ref{den2}.a compares the differential capacitance computed with Eq.~(\ref{varel}) in the vanishing surface charge limit with experimental data obtained for several types of monovalent electrolytes at various concentrations (for details see Ref.~\cite{prlnetz}  where the data were taken from). We also report in this figure the prediction of various formalisms. As stressed in Ref.~\cite{prlnetz}, the PB result largely overestimates the experimental data. Furthermore, the result of the modified PB (MPB) equation (see Ref.~\cite{pre}) that can exclusively take into account the ionic depletion effect brings a very small correction to the PB result. However, the EDPB result that additionally  contains the surface depletion effect of solvent molecules exhibits a good agreement with the experimental data. We finally note that in the vanishing surface charge limit considered in this part, the DPB equation yields the same result as the PB one (see black squares in Fig.~\ref{den2}.a).

The physics of the EDPB prediction for the capacitance can be understood within the restricted self-consistent scheme of Eq.~(\ref{phires}), where the differential capacitance in Eq.~(\ref{cap1}) takes in the limit $\sigma_s\to0$ the simple form
\be\label{cap2}
C_d=\frac{\e_w\kappa_c}{1+\kappa_c(d-h)+\e_w\kappa_c h}.
\ee
We note that the prediction of this equation reported in Fig.~\ref{den2} (red squares) fits very well the numerical result of the EDPB equation. The inverse capacitance of Eq.~(\ref{cap2}) is composed of three parts. The first contribution from the diffuse layer is the inverse GC capacitance $C_{d1}^{-1}=(\e_w\kappa_c)^{-1}$ corresponding to the PB result in Fig.~\ref{den2}.a. The second part $C_{d2}^{-1}=(d-h)/\e_w$ associated with the ionic depletion layer is shown to drop the differential capacitance to the MPB curve. Finally, the third contribution from the solvent depletion layer $C_{d3}^{-1}=h$ characterized by the dielectric screening deficiency brings the most important correction to the total capacitance by dropping the latter to the correct order of magnitude.

In order to estimate the order of magnitude of the dipolar depletion length $h$, we will compute the asymptotic limit of $\te(z)$ far from the interface, where the dipolar potentials in Eqs.~(\ref{funu}) and~(\ref{funt}) become weak, by expanding Eq.~(\ref{per2}) in $U_d(z)$, $T_d(z)$, and $p_0\phi'_0(z)$. Furthermore, we note that in the limit $\e_m=0$, the dipolar potentials are given by the closed form expressions
\bea
\label{ud}
U_d(z)&=&\ell_wp_0^2\frac{1+2\kappa_cz}{16z^3}e^{-2\kappa_cz}\\
\label{td}
T_d(z)&=&\ell_wp_0^2\frac{1+2\kappa_cz(1+2\kappa_cz)}{16z^3}e^{-2\kappa_cz}.
\eea
Renormalizing all lengths by the length scale $l_d=(\ell_wp_0^2/10)^{1/3}$ according to $\bar\kappa_c=\kappa_cl_d$ and $\bar z=z/l_d$, and taking into account that $\e_w\gg1$, the asymptotic form of Eq.~(\ref{per2}) far from the dielectric interface reads
\be
\label{diel4}
\frac{\te(z)}{\e_w}\approx1-\frac{1+2\bar\kappa_c\bar z+3\bar\kappa_c^2\bar z^2/2}{\bar z^3}e^{-2\bar\kappa_c\bar z}.
\ee
We now note that for a bulk permittivity $\e_w=71$ and ionic concentration $\rho_{bi}=10^{-1}$ M, one has $\kappa_cl_d=8.10^{-2}$. Hence, in the regime $z\sim l_d=0.9$ {\AA}, the terms in Eq.~(\ref{diel4}) that depend on the screening length become negligible. This simple calculation fixes $l_d$ as the characteristic length over which the local permittivity tends to its bulk value according to an inverse cubic power law, i.e. $\te(z)/\e_w\approx1-l_d^3/z^3$. We note that an inverse cubic law for the dielectric permittivity profile was derived in Ref.~\cite{hansenepl} in the strict limit of a single dipole (i.e. $\e_w=1$) and without salt. In the dilute salt limit $\kappa_c\to 0$, one can actually extend the estimation of $h$ to finite values of $\e_m$ by noting that the dipolar potentials possed the close form expression $U_d(z)=T_d(z)=\ell_wp_0^2\Delta_0/(16z^3)$, where $\Delta_0=(\e_w-\e_m)/(\e_w+\e_m)$. Following the same steps as above, one obtains for the characteristic dipolar depletion length the more general expression $l_d=(\Delta_0\ell_wp_0^2/10)^{1/3}$.  Hence, for biological solvent concentrations and dilute electrolytes with bulk density $\rho_{bi}\lesssim 0.1$ M, the dielectric screening is mainly responsible for the decay of the image dipole interactions and solely determines the region over which a reduced dielectric permittivity is observed. For larger ion concentrations, Eq.~(\ref{diel4}) shows that the screening of image dipole interactions by surrounding ions positively adds to the dielectric screening of these forces.

We display in the inset of Fig.~\ref{den2}.b the evolution of $h$ as a function of $\e_m$ together with $l_d$, while the main plot shows $C_d$ for various values of the membrane permittivity from $\e_m=1$ to $\e_m=\e_w$. One notices that within the range $1\leq\e_m\leq60$, $h$ exhibits a slow linear decrease with increasing $\e_m$ while $C_d$ remains within the same order of magnitude as the experimental capacitance data. However, with an increase of $\e_m$ from 60 to $\e_w$, the dipolar depletion length quickly drops to zero and consequently, $C_d$ approaches the PB result. One also sees in the inset that although $l_d$ is slightly higher than the dipolar depletion length $h$, it can reproduce the correct trend of the latter as a function of $\e_m$. These observations  suggest that image dipole interactions are mainly responsible for the low values of the experimental capacitance data in Fig.~\ref{den2}.a. We emphasize that this result is in agreement with the experimental observation of a strong reduction of the double layer capacitance with increasing surface hydrophobicity~\cite{exp}.

\begin{figure}
(a)\includegraphics[width=0.95\linewidth]{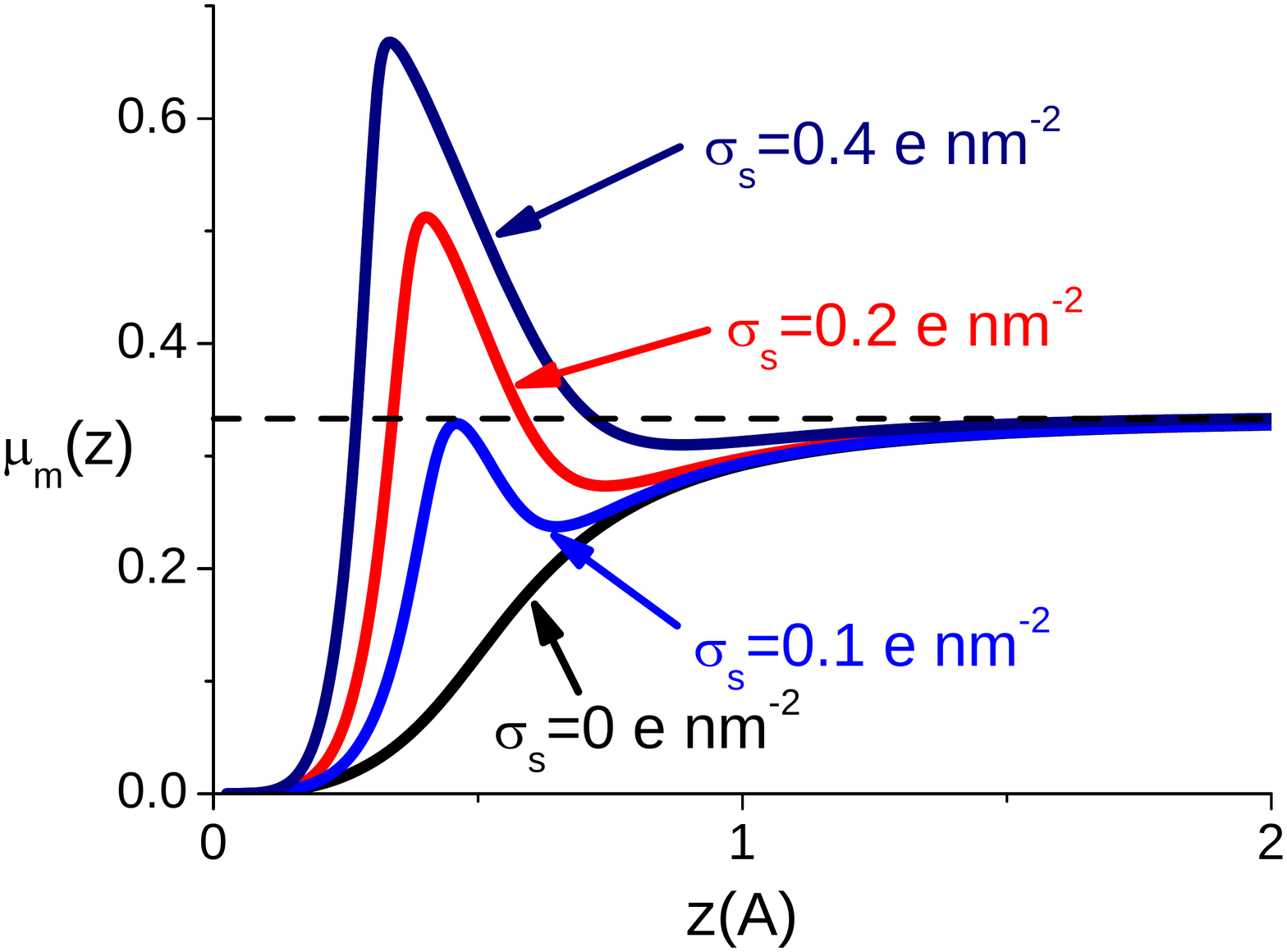}
\caption{(Color online)  (a) Dipolar orientation profile evaluated for various $\sigma_s$ and the same model parameters as in Fig.~\ref{den2}. The black dashed reference line marks the freely rotating dipole case $\mu_m(z)=1/3$.}
\label{den3}
\end{figure}

In addition to the dipolar depletion effect, the orientation is also expected to play some role in the form of the interfacial dielectric permittivity profile. The measure of the dipolar orientation is defined in the literature as $\mu_m(z)=\lan p_z^2\ran/[p_0^2\rho_d(z)]$, where $\lan p_z^2 \ran=\int\frac{\mathrm{d}\bom}{4\pi}\bar\rho_d(z,\bom)p_z^2$. The function $\mu_m$ was studied in Ref.~\cite{Kanduc} for multipolar ions and it was found invariably below the free dipole value 1/3 in the SC limit (i.e. dipolar alignment parallel to the wall) and above this value in the WC limit (alignment along the electrostatic field). We show in Fig.~\ref{den3} that for a neutral interface, one has $\mu_m(z)<1/3$ as in the SC limit, i.e. the solvent molecules exhibit a tendency to align parallel to the wall over a distance $\approx 2$ {\AA}, that is, until image dipole forces vanish. We now define an effective dielectric permittivity function of  the form $\tilde\e_{eff}(z)=1+4\pi\ell_Bp_0^2\rho_d(z)/3$ that solely accounts for the dipolar depletion. The comparison of $\tilde\e_{eff}(z)$ in Fig.~\ref{den1}.a with $\tilde\e(z)$ shows that the main effect of the dipolar alignment close to the interface is a slight reduction of the local dielectric permittivity. However, it is seen that this effect is largely dominated by the solvent depletion.

In the presence of a finite surface charge, Fig.~(\ref{den3}) shows that interestingly, the function $\mu_m(z)$ exhibits a non-monotonous behavior. Namely, in the close vicinity of the surface, strong image dipole effects lead to a net dipolar alignment parallel to the wall. However, above a characteristic distance from the dielectric interface where the surface charge induced electric field $p_0\phi'_0(z)$ dominates the image dipole potential, $\mu(z)$ exceeds 1/3 and reaches a peak where the dipoles exhibit the maximum tendency to align in the direction of the field, i.e. perpendicular to the dielectric wall. With increasing distance, one notices a reversal of this behavior where the image dipole potential dominates for a second time the electrostatic field. As a result, the solvent molecules exhibit again some tendency to align again parallel to the interface, but this regime gradually disappears with increasing surface charge.

\section{Conclusions}
We have considered the dielectric discontinuity effects on the differential capacitance of low dielectric substrates. To this aim,  we derived an extended DPB equation that can explicitly account for the interactions of solvent molecules with their electrostatic image. Within this approach, we showed that the overestimation of the experimental data by the GC capacitance is due to the inability of the latter to account for the solvent depletion effect driven by image dipole interactions. The prediction of the EDPB equation for the differential capacitance of monovalent electrolytes was compared with experimental data and good agreement was found.

The EDPB formalism is a first order theoretical approach in the explicit modeling of solvent molecules beyond the MF level, and it has its limitations. Excluded volume~\cite{jcp}  and non-local dielectric effects that lead in MD simulations to interfacial structure formation~\cite{prlnetz} are absent. The theory could be extended by using more general trial kernels, but the analytical solution of the Debye-Huckel equation with a local dielectric permittivity and screening parameter is still an open problem. Furthermore, the present formalism does not include multipolar moments, which are known to enhance the interfacial dielectric exclusion\cite{Kanduc}. Hence, multipolar contributions are expected to further lower the capacitance curves in Fig.~\ref{den2}.
\acknowledgements This work has been supported in part by The Academy of Finland through its COMP CoE and NanoFluid grants.

\end{document}